\documentclass[conference]{IEEEtran}
\IEEEoverridecommandlockouts
\usepackage{cite}
\usepackage{amsmath,amssymb,amsfonts}
\usepackage{algorithmic}
\usepackage{graphicx}
\usepackage{multirow}
\usepackage{rotating}
\usepackage{textcomp}
\usepackage[table]{xcolor}
\usepackage{makecell}
\usepackage{xcolor}
\usepackage{hyperref, enumitem}
\usepackage{tikz}
\usetikzlibrary{positioning}
\usepackage{booktabs,multirow}

\def\BibTeX{{\rm B\kern-.05em{\sc i\kern-.025em b}\kern-.08em
    T\kern-.1667em\lower.7ex\hbox{E}\kern-.125emX}}

\begin{document}

\title{Code Division Modulation Layers \\ Against Forgetting and Inference \\ in Continual Gait Identification
\thanks{
We thank the partnership and collaboration with the São Paulo Research Foundation (Fapesp) Horus project, Grant \#2023/12865-8.
This manuscript reflects only the authors’ views and opinions, neither EU nor the European Commission nor FAPESP can be considered responsible for them. All of the authors have revised the document and confirm the findings.}
}

\author{
\IEEEauthorblockN{Simone Milani}
\IEEEauthorblockA{\textit{Department of Information Engineering,  University of Padova, Padova, Italy} \\
\texttt{simone.milani@dei.unipd.it}}
}

\maketitle

\begin{abstract}
Continual learning (CL) has been recently employed in biometric identification systems thanks to its ability to integrate new knowledge within a pre-trained model and to the possibility of reducing the computational cost of training. Unfortunately, such approaches pose new challenges both in terms of final accuracy and  privacy guarantees since a progressive fine-tuning of the model on small subsets expose them to catastrophic forgetting and successful inference attacks. 

This paper evaluates the efficiency of code division modulation layers (CDML) on a gait identification system which has been trained following a continual learning policy. The proposed approach preserves accuracy on all the tasks while mitigating membership inference attacks at the same time. Moreover, the impact of retransmission is minimized since replaying data is not necessary. 
\end{abstract}


\section{Introduction and related works} \label{sec:introduction}

\begin{figure}[t]
  \centering
  \centerline{\includegraphics[width=\columnwidth]{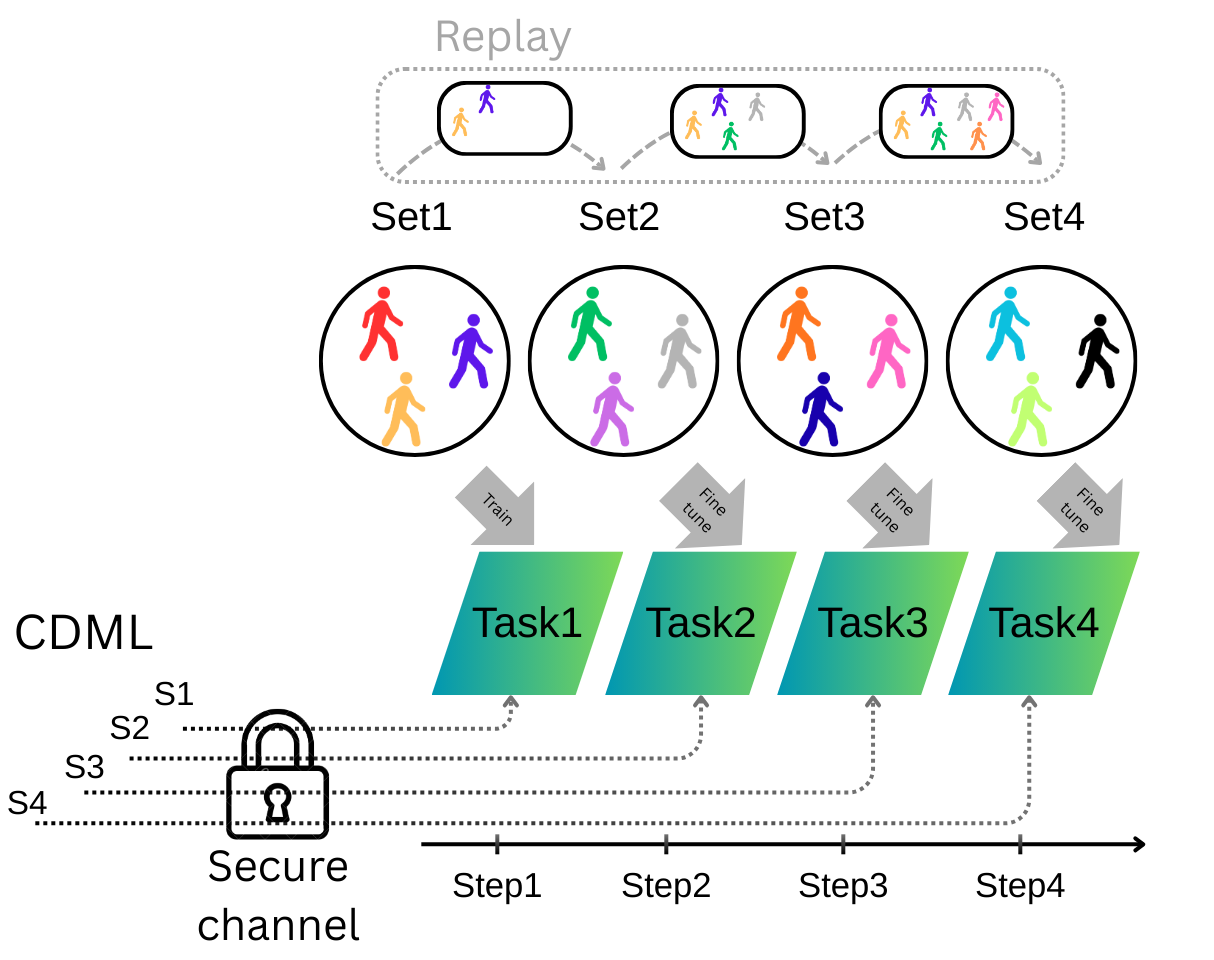}}
\caption{Continual Learning scheme for gait-based identification. The block diagram reports the eventual replay strategy and the use of CDML information to mitigate catastrophic forgetting.}
\label{fig:cl}
\end{figure}

Learning-based biometric systems have recently focused on studying continual learning (CL) approaches \cite{zhao2020continual}  in order to integrate new data, individuals or datasets into a pre-trained model (see the block diagram in Fig.~\ref{fig:cl}). This interest  arises from the computational limitations posed to retraining a model from scratch whenever an update is needed \cite{RAHMAN2025102044} and from privacy policies that may prevent biometric data sharing across different stages or terminals \cite{KATKAR202636}. Moreover, behavioral biometrics are often characterized by non-stationary and time-varying templates depending on the acquisition contexts and conditions of the subjects \cite{baig}.  Indeed, updating the model step-by-step (or task-by-task) permits protecting sensitive information during updates and constraining the computational load, and often, these solutions are implemented in a federated \cite{10148063} or differential privacy set-up \cite{hassandp}.

Unfortunately, CL solutions enable flexibility but pose new problems and challenges. One of the major ones is \emph{catastrophic forgetting}, i.e., the information related to the first learning stages is lost after a few stages of training, and hence, the accuracy of the model reduces on the first classes or modalities. In order to mitigate this, several solutions were proposed ranging from replaying previous data \cite{math13142257}, using coarse-to-fine knowledge distillation \cite{Camuffo2023ContinualLF}, adopting modulation layers that remap the learned features \cite{film,loo2020combining} or injecting data correlated with previous information and created with generative models \cite{liang2024dddr}, to mention some of them.

All these solutions mitigate the performance decrement, but produce a decrement in the privacy level since some sensitive information needs to be shared across tasks (see the diagram in Fig.~\ref{fig:cl}).

Moreover, a second problem concerns the possibility of non-disclosing the partial datasets of each task by sharing the model only. Since the size of the trained network is dimensioned on the final overall task, the amount of data used to train the network at each task results limited. This exposes the generated network to possible Membership Inference Attacks (MIA) or Identity Inference Attacks (IIA). These attacks denote all those strategies that aims at inferring whether a given sample (MIA) or class/subject (IIA) was used in the training set of the probed model \cite{NIU2024404,Nasr2018ComprehensivePA}. Such strategies have been designed for classifiers \cite{9833649,Yeom2017PrivacyRI}, generative models \cite{CAVASIN2024184}, and biometric recognition systems \cite{11159206}. Indeed, success probabilities of attacks increase as the cardinality of training datasets reduces with respect to the size of the model \cite{9833649,CAVASIN2024184}.

The current paper proposes a class-incremental continual learning strategy for gait identification that aims at solving both catastrophic forgetting and exposure to inference attacks by including a Code Division Modulating Layer (CDML) on the feature space that embeds the input signal to be classified. The approach is applied to the gait classifier reported in \cite{zou2020deep} and deployed in a continual learning set-up where users are divided across multiple tasks. Experimental results show that the final accuracy in classification is maximized while preventing inference in a black box scenario.

The main novelties introduced by the paper can be summarized as follows.
\begin{itemize}
\item The paper analyzes the impact of different countermeasures against catastrophic forgetting in Continual Learning from the privacy perspective by measuring the success probability of membership inference attacks.
\item A novel Code Division Modulating Layer (CDML) is introduced in order to mitigate both the accuracy loss and the possibility of membership inference on a standard gait identification classifier. The proposed solution does not necessarily require the transmission of replay data, thus minimizing the transmission and storage needs.
\item The different methods are then combined together to identify the most effective solution.
\item An experimental validation of both classification accuracy and privacy level is performed for each solution. Results show that the proposed scheme improves identification performance with negligible impact on robustness and complexity.
\item The approach enables granting different privacy levels to different users according to which modulating codes are shared.
\end{itemize}

These aspects are detailed in the following subsections.

\section{The proposed approach}

\begin{figure}[t]
  \centering
  \centerline{\includegraphics[width=\columnwidth]{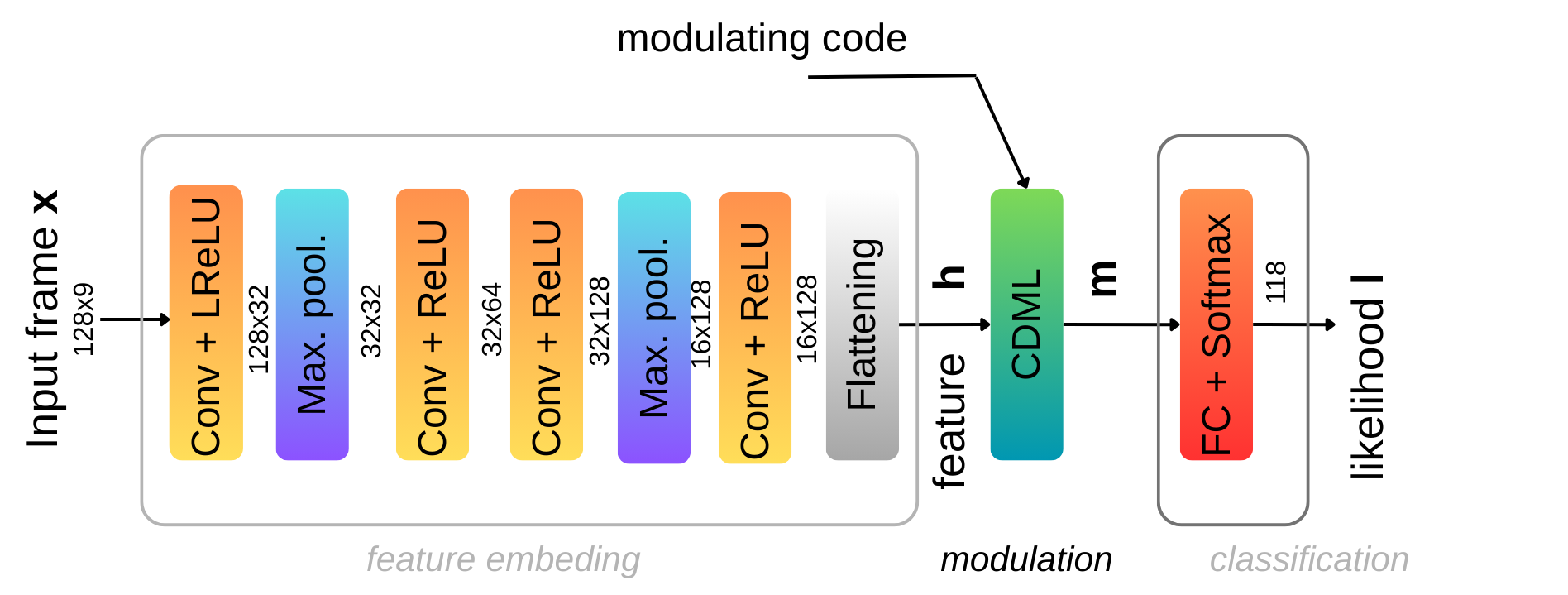}}
\caption{Network structure from \cite{zou2020deep} with the additional CDML layer.}
\label{fig:net}
\end{figure}

The starting point of the proposed approach is the CNN-based  gait identification model in \cite{zou2020deep}, which was chosen since results can be easily generalized and its structure can be well adapted to low-power devices with minimal quality loss as shown in \cite{10810715}. The general structure is reported in Fig.~\ref{fig:net} (gray circled blocks), together with the additional CDML layer proposed by this work. The network is made by 1D Convolutional layers, followed by ReLU activations and MaxPooling layers. The input is an array of $128$ samples with $9$ dimensions corresponding to the Inertial Measurement Unit (IMU) of a cellphone (more details in \cite{zou2020deep}).

Whenever the training set is decomposed into multiple subsets associated with a CL task, catastrophic forgetting occurs and therefore, mitigation strategies may involve replaying previous data or using prototypes.

As for replay solutions, the current paper refers to the work \cite{math13142257} where a percentage of samples from Tasks $1, \ldots, k-1$ are used to train the network for Task $k$.
Another possible countermeasure is the inclusion of a Feature-wise Linear Modulation (FiLM) layer \cite{film} that adapts the features $\mathbf{h}=f(\mathbf{x}$ from the embedding layer using a set of weights $\gamma_k$ and biases $\beta_k$,
where both $\gamma_k$ and $\beta_k$ are computed during the incremental training at step/task $k$. This approach was first introduced in the general field of Continual Learning in \cite{film}, and it has been recently used for adaptation in gait identification \cite{Wang2025GaitAdaptCL}. The solution allows avoiding replaying data and preserves previously-learned features thus keeping the detection accuracy high. Unfortunately, it  requires the transmission of parameters $\gamma_k,\beta_k$ to the next classifier. Indeed, diversified privacy levels are enabled since each terminal can be provided with customizable detection capabilities depending on which set of $\gamma_k$, $\beta_k$  is used.

The proposed approach implements a code division modulation on features $\mathbf{h}$ by generating a set of uncorrelated sequences $\mathbf{s}_k$, $k=1, \ldots, N$, where $s_{k,i} \in \{-1,+1\}$ and $\mathbf{s}_k \cdot \mathbf{s}_h \approx 0$ $\forall k \neq h$. Sequences are created using a pseudo-random generator followed by a quantization stage. The seed is transmitted to the final enabled users through a secure channel (not publicly-distributed), who can re-generate the full sequence. Once the sequence is generated, it stays fixed and is not altered during the training process.

\begin{table}[t]
\centering
\caption{Comparison of a uniform and Gaussian distribution and different generators using statistical tests.}
\label{tab:PRGComp}
\footnotesize
\begin{tabular}{|l|r|r|r|r|}
\hline
\textbf{Dist/Gen.} & \textbf{CS} & \textbf{PCC} & \textbf{KS Test} & \textbf{Run Test} \\ \hline
Uniform Dist            & -2.09e-07 & -2.13e-07 & 0.506 & -0.0195 \\ \hline
Gaussian Dist           & 5.28e-03 & 1.43e-06 & 0.186 & 0.0115 \\ \hline
Hénon Map               & 1.27e-03 & 7.32e-05 & 0.365 & 0.485 \\ \hline
Logistic Map            & 7.97e-03 & 2.52e-05 & 0.198 & 4.326 \\ \hline
Tent Map                & 8.43e-03 & 1.30e-05 & 0.144 & 2.496 \\ \hline
LCG                     & -1.46e-05 & -1.38e-05 & 0.505 & -0.005 \\ \hline
\end{tabular}
\end{table}

The modulated features 
\begin{equation}
\mathbf{m} = \mathbf{s}_k \odot \mathbf{h} 
\end{equation}
\noindent are then processed by the classification layer and mapped into a likelihood array $\mathbf{l}=g(\mathbf{m})$, which will provide the outcome of the classification.

\begin{figure*}[t]
  \centering
  \begin{minipage}[c]{0.66\columnwidth}
  \includegraphics[width=\columnwidth]{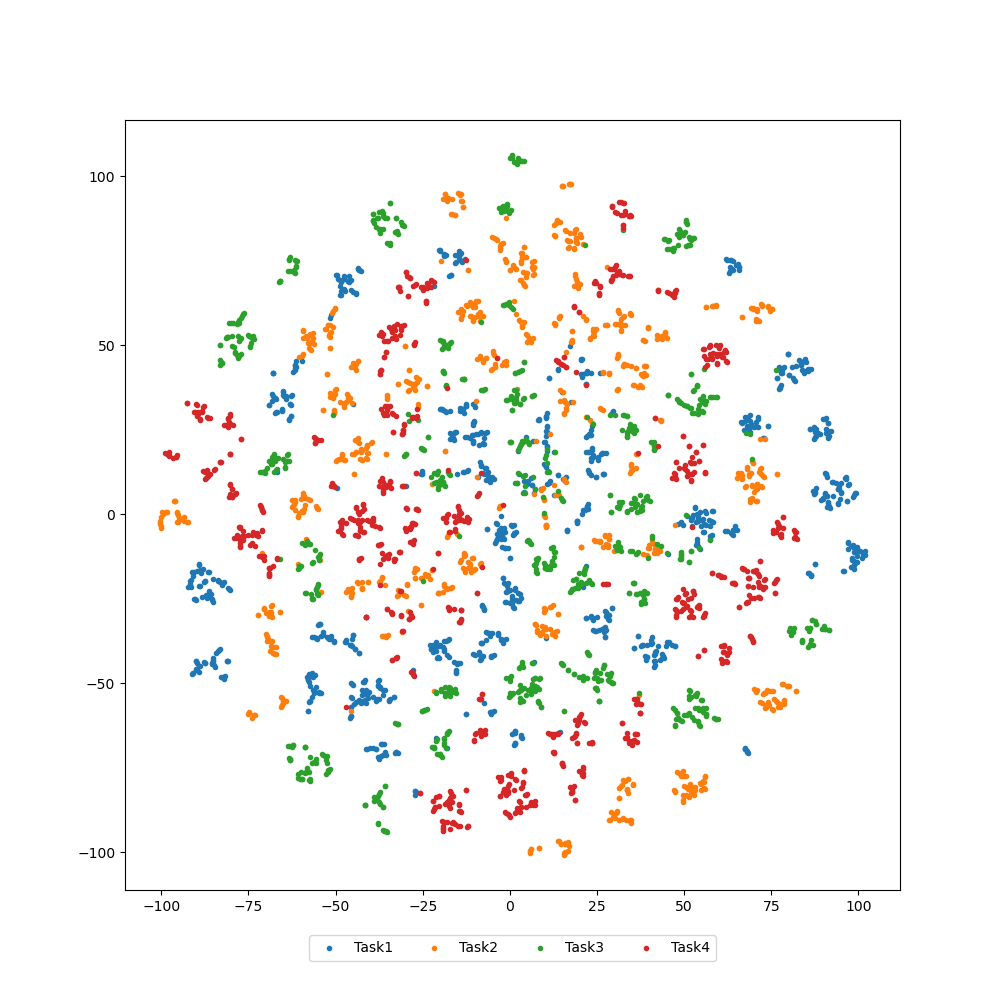} \\
  \centerline{Std. \cite{math13142257}}
  \end{minipage} \hfil
   \begin{minipage}[c]{0.66\columnwidth}
  \includegraphics[width=\columnwidth]{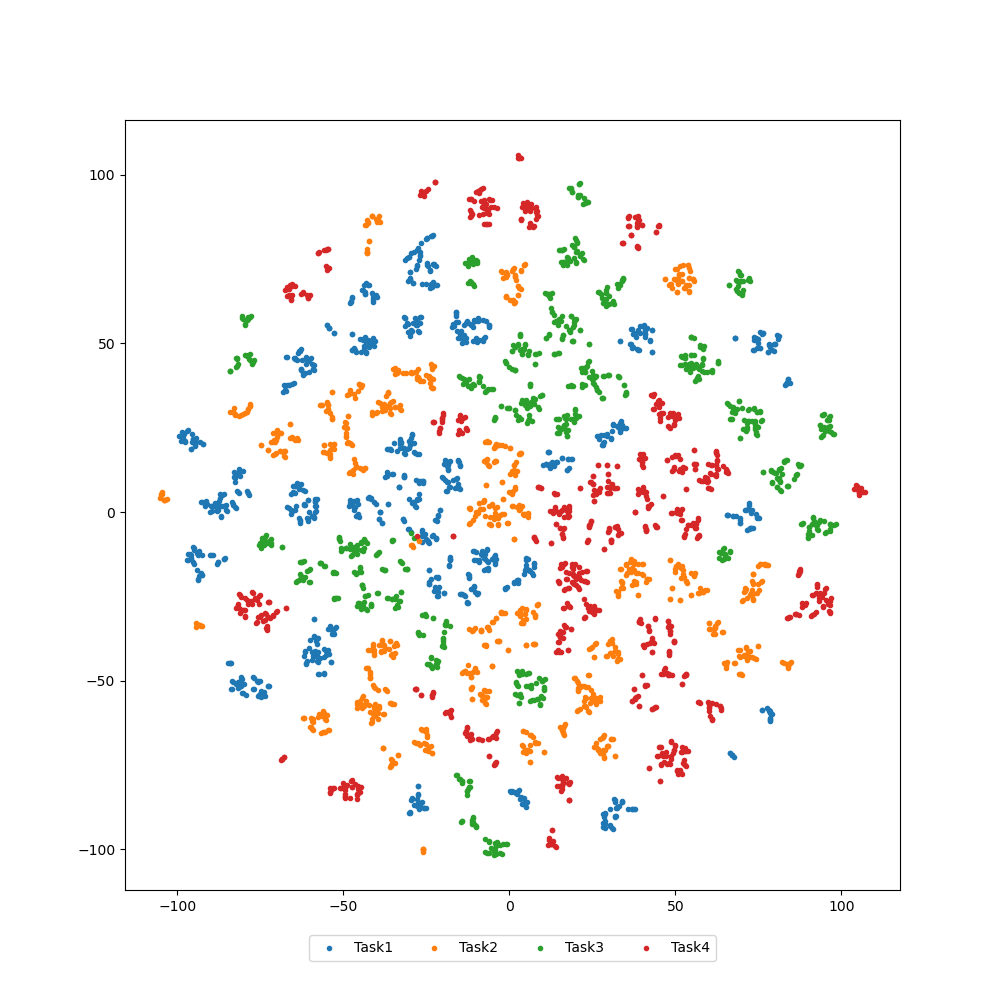}
  \\
  \centerline{Adapt \cite{Wang2025GaitAdaptCL,film}}
  \end{minipage} \hfil
   \begin{minipage}[c]{.66\columnwidth}
  \includegraphics[width=\columnwidth]{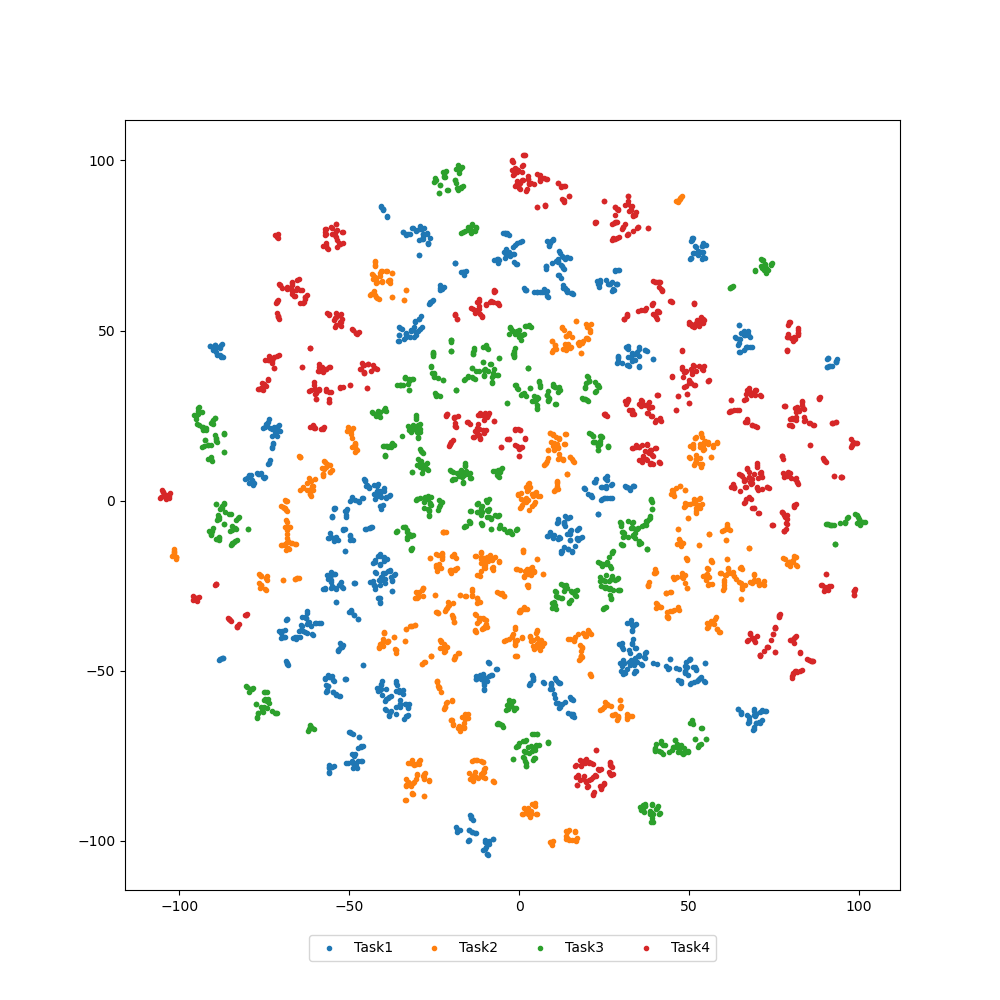}
  \\
  \centerline{CDML (proposed)}
  \end{minipage}
\caption{Plots of t-SNE projections for the latent representation associated to different CL solutions. Different colors are associated to samples belonging to different tasks.}
\label{fig:tsne}
\end{figure*}

This processing step is analogous to Code Division Multiple Access (CDMA) systems in communications and allows embedding data from different tasks into uncorrelated feature spaces. In this way, even if the input samples (associated to different IDs) are mapped into a common latent region (because of signal similarities), feature modulation separates them into diverse uncorrelated areas thus enabling a correct classification.

This allows protecting training data from inference attacks as described in the following section.

From these premises, the crucial requirement for the selected random noise generator is  the ability to create uncorrelated modulation sequences. 
For this purpose, two metrics have been used to measure correlation (\textit{cosine similarity} (\textit{CS}) and \textit{Pearson correlation coefficient} (\textit{PCC})) between different sequences generated by the considered algorithms.  
Two other measures assess the randomness level of the generated sequence: the (\textit{Kolmogorov-Smirnov test (\textit{KS test}) and the \textit{Wald-Wolfowitz runs test} (\textit{run test})) \cite{randomness}.}
The analysis has been performed over four different pseudorandom functions: \textit{Hénon noise} \cite{Henon}, \textit{Logistic map} \cite{Lostic_map}, \textit{Tent map} \cite{Tent_map} and \textit{Linear Congruential Generator} (\textit{LCG}) \cite{LCG_def}.
Each function output  has been normalised into the range $[-1,1]$ and results are reported in Table~\ref{tab:PRGComp}.

In order to evaluate the impact of the modulation scheme on the final latent representation, t-SNE algorithm was applied on the feature arrays associated to test data.
In this evaluation, testing considered a standard replay (\texttt{Std.}), learned-adaptation (\texttt{Adapt}), and the proposed strategy. Experimental results are reported in Fig.~\ref{fig:tsne}, and it is possible to notice that CDML achieves a better homogeneity in region organization.

\section{Inference attacks}

In literature, different MIA strategies have been proposed during the last years. This work focuses on black box attacks where the inner structure of the network and weight values are not known by the attacker. Indeed, the attacker wants to verify whether a given couple $(\mathbf{x}_i,\mbox{id}_i)$ belongs to the training set or not, where $\mathbf{x}_i$ is the input sample and $\mbox{id}_i$ is the corresponding ground truth label. 

In this work, a confidence-based MIA attack (based on LiRA) was implemented \cite{9833649,Yeom2017PrivacyRI}, where likelihoods $\mathbf{l} = g(\mathbf{m})$ are computed and the value $\mathbf{l}[\mbox{id}_i]$ is compared to a threshold. If the value is higher, it is possible to assume that the model displays a high confidence in the classification results and therefore, the sample is likely to belong to the training set.

Better strategies have been proposed in literature (e.g., using gradients, shadow models, etc.), but experimental results show that on the networks in \cite{zou2020deep} this strategy is sufficiently successful despite the limited size of the models.

\section{Experimental set-up and results}
\begin{table}[t]
    \centering
        \caption{Subset partitioning across different tasks}
    \label{tab:tasks}
    \begin{tabular}{|c|c|c|c|c|}
    \hline
    & \footnotesize Task 1   &  \footnotesize Task 2  &  \footnotesize Task 3 & \footnotesize  Task 4 \\ 
    \cline{2-5}
    & \footnotesize Set 1   &  \footnotesize Set  2  &  \footnotesize Set 3 & \footnotesize  Set 4 \\ \hline 
 \footnotesize Subjects & $0-29$ & $30-59$ & $60-89$ & $90-118$     \\ 
 \hline
    \end{tabular}
\end{table}
\subsection{Setting}
In the experiments, the training set is divided into $4$ subsets including separate sets of subjects/users as reported in Table~\ref{tab:tasks}. Subjects from dataset $1$ of \cite{zou2020deep} are divided into groups of 30 persons and associated to $4$ different tasks. Input data are made of $6$ or $9$ dimensional arrays of $128$ samples, which were segmented from longer sequences according to gait cycles. Dataset acquisition and preparation are described in \cite{zou2020deep}. At first, the network is trained on set1, and then fine-tuned on the following subsets for $400$ epochs with an exponentially decaying learning rate (starting at $0.001$ with factor $0.9$). 

\begin{table*}[t]
\centering
\setlength\tabcolsep{4pt}
\renewcommand\arraystretch{0.9}
\caption{AUC (\%) / EER(\%) for MIA attacks on different configurations. Performances are divided into steps, where the data related to each task are reported. Best and second best results are highlighted in blue.}\label{tab:attack}

\begin{tabular}{|c|c||c||c|c||c|c|c||c|c|c|c|}
\hline
&  \footnotesize \bf Replay  & \multicolumn{1}{c||}{\footnotesize \bf Step 1}  & \multicolumn{2}{c||}{\footnotesize \bf Step 2} & \multicolumn{3}{c||}{\footnotesize \bf Step 3} & \multicolumn{4}{c|}{\footnotesize \bf Step 4} \\
\hline
& &
\footnotesize Task$1$ & \footnotesize Task$1$ & \footnotesize Task$2$
& \footnotesize Task$1$ & \footnotesize Task$2$ & \footnotesize Task$3$ & \footnotesize Task$1$ & \footnotesize Task$2$ &\footnotesize Task$3$ & \footnotesize Task$4$ \\
\hline
\multirow{3}{*}{\begin{turn}{90} \footnotesize \bf Std. \end{turn} } 
 & \footnotesize  0 \%  & \footnotesize $56.8$/$31.5$  & \footnotesize $56.0$/$18.4$  & \footnotesize $50.5$/$28.2$  & \footnotesize $56.9$/$24.1$  & \footnotesize $50.5$/$45.2$  & \footnotesize $54.3$/$47.8$  & \footnotesize $57.0$/$21.2$  & \footnotesize $57.8$/$45.0$  & \footnotesize $53.9$/$38.8$  & \footnotesize $54.9$/$34.2$ \\
\cline{2-12} 
 & \footnotesize   10\%  & \footnotesize $57.3$/$31.5$  & \footnotesize $55.6$/$19.4$  & \footnotesize $49.7$/$13.7$  & \footnotesize $56.3$/$26.3$  & \footnotesize $52.7$/$39.8$  & \footnotesize $51.1$/$48.1$  & \footnotesize $55.2$/$21.5$  & \footnotesize $60.8$/$45.0$  & \footnotesize $54.1$/$47.0$  & \footnotesize $53.5$/$43.0$ \\
\cline{2-12} 
 & \footnotesize   30\%  & \footnotesize $56.0$/$31.5$  & \footnotesize $55.5$/$26.4$  & \footnotesize $51.8$/$37.6$  & \footnotesize $56.5$/$30.9$  & \footnotesize $53.6$/$47.0$  & \footnotesize $54.3$/$43.3$  & \footnotesize $54.7$/$37.0$  & \footnotesize $57.4$/$42.6$  & \footnotesize $52.2$/$47.1$  & \footnotesize $51.4$/$48.2$ \\
\hline\hline
\multirow{3}{*}{\begin{turn}{90} 
\footnotesize \bf Adapt  \end{turn} }
 & \footnotesize  0 \%  & \footnotesize $56.9$/$34.7$  & \footnotesize $54.9$/$46.6$  & \footnotesize $48.0$/$33.7$  & \footnotesize $54.0$/$48.3$  & \footnotesize $51.9$/$30.1$  & \footnotesize $52.4$/$35.1$  & \footnotesize $54.9$/$46.8$  & \footnotesize $50.5$/$32.7$  & \footnotesize $50.6$/$32.6$  & \footnotesize $50.8$/$33.4$ \\
\cline{2-12} 
 & \footnotesize   10\%  & \footnotesize $57.4$/$37.0$  & \footnotesize $57.4$/$45.0$  & \footnotesize $49.9$/$27.1$  & \footnotesize $55.8$/$45.9$  & \footnotesize $49.4$/$26.0$  & \footnotesize $54.6$/$35.7$  & \footnotesize $56.5$/$45.4$  & \footnotesize $49.6$/$41.1$  & \footnotesize $51.0$/$43.5$  & \footnotesize $52.8$/$44.0$ \\
\cline{2-12} 
 & \footnotesize   30\%  & \footnotesize $57.2$/$32.9$  & \footnotesize $54.5$/$47.6$  & \footnotesize $49.8$/$29.1$  & \footnotesize $54.4$/$47.9$  & \footnotesize $48.7$/$35.3$  & \footnotesize \cellcolor{blue!30} $48.8$/$49.7$  & \footnotesize $53.3$/$48.4$  & \footnotesize  \cellcolor{blue!10}  $48.5$/$48.4$  & \footnotesize \cellcolor{blue!30} $50.1$/$49.8$  & \footnotesize \cellcolor{blue!30} $49.8$/$49.5$ \\
\hline \hline
\multirow{3}{*}{\begin{turn}{90} \footnotesize \bf CDML \end{turn} } 
  & \footnotesize  0 \%  & \footnotesize \cellcolor{blue!10}   $49.0$/$49.6$  & \footnotesize \cellcolor{blue!30} $50.1$/$49.6$  & \footnotesize $49.0$/$48.5$  & \footnotesize \cellcolor{blue!30} $49.4$/$49.3$  & \footnotesize \cellcolor{blue!10}  $50.3$/$49.6$  & \footnotesize \cellcolor{blue!10}  $51.3$/$48.8$  & \footnotesize $49.8$/$48.8$  & \footnotesize \cellcolor{blue!30} $49.5$/$49.1$  & \footnotesize $52.4$/$48.2$  & \footnotesize $51.2$/$48.7$ \\
\cline{2-12} 
 & \footnotesize   10\%  & \footnotesize $48.5$/$48.5$  & \footnotesize $47.6$/$49.0$  & \footnotesize \cellcolor{blue!10}  $50.6$/$48.5$  & \footnotesize $48.4$/$48.3$  & \footnotesize $50.8$/$49.6$  & \footnotesize $49.4$/$48.5$  & \footnotesize \cellcolor{blue!30} $50.8$/$49.6$  & \footnotesize $51.4$/$45.6$  & \footnotesize $51.4$/$48.6$  & \footnotesize \cellcolor{blue!10}  $50.7$/$49.1$ \\
\cline{2-12} 
 & \footnotesize   30\%  & \cellcolor{blue!30} \footnotesize $49.8$/$49.7$  & \footnotesize \cellcolor{blue!10}  $48.4$/$49.0$  & \footnotesize \cellcolor{blue!30} $49.0$/$48.8$  & \footnotesize \cellcolor{blue!10}  $51.2$/$48.7$  & \footnotesize \cellcolor{blue!30} $50.5$/$49.9$  & \footnotesize $54.0$/$47.8$  & \footnotesize \cellcolor{blue!10}  $52.1$/$49.2$  & \footnotesize $53.0$/$48.0$  & \footnotesize \cellcolor{blue!10} $50.6$/$49.1$  & \footnotesize $47.7$/$48.5$ \\
\hline
\end{tabular}

\end{table*}

The proposed CDML strategy is compared with state-of-the-art replay strategies \cite{math13142257}, where a percentage of previous datasets is transmitted to the following tasks. The percentage of re-processed data may be tuned according to the average accuracy.

In addition, the experimental comparison also tested an adapter module (following the same strategy reported in \cite{Wang2025GaitAdaptCL}) based on a FiLM layer that is tuned for each task in order to remap features. The paper refers to this approach as \texttt{Adapt}.

The CDML setting assumes that each task $k$ is associated with a modulating code $\mathbf{s}_k$ which is transmitted to the next stages. In order to correctly-classify a sample, users must know the correct modulation code $\mathbf{s}_k$, otherwise they can apply a random guessing or use a constant modulation.


Performance is evaluated measuring classification accuracy (to evaluate the impact of forgetting) and of Area Under Curve (AUC) and the Equal Error Rate (EER) from ROC curves whenever considering the efficiency of a MIA attack (AUC needs to be as close as possible to $50$ \%, while EER is to be maximized).

\subsection{Results}
\begin{figure}[t]
  \centering
  \centerline{\includegraphics[width=\columnwidth]{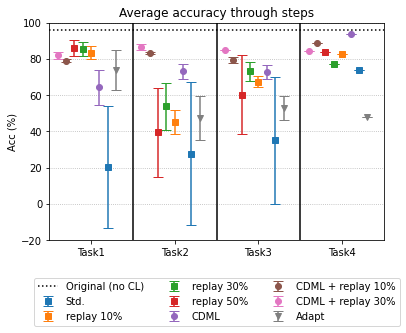}}
\caption{Average classification accuracy for different CL methods across the different steps. Vertical bars denote accuracy variance across different training sets. Dashed line relates to architecture in \cite{zou2020deep}, while solid lines are related to the proposed CDML-based approach and Adapt.}
\label{fig:acc}
\end{figure}

At first, the impact of catastrophic forgetting was evaluated  across the different steps for different strategies. In the experimental validation, tests compared a standard fine-tuning (labeled \texttt{Std}), strengthened by replay with different percentages of repetition (see Fig.~\ref{fig:acc}). The CDML approach permits mitigating the accuracy loss without requiring data retransmission and reprocessing since accuracy is comparable to the one obtained with replay methods. 
As for \texttt{Adapt} approach, similar consideration can be done although in this case the average performance is lower since the number of trainable parameters increases, thus requiring more data and computational complexity.

Note also that minimal replay can be adopted for CDML as well, thus maximizing the final performance without requiring a significant retransmission ($10$ \%). Another advantage of CDML lies in reduced accuracy variance across the different tasks (vertical bars of the graph): the use of CDML layer permits smoothing the accuracy throughout the different subsets, while in the standard set-up values depend on the processing order. 

Tests also verified that the robustness to inference attacks is compromised by replaying. Table~\ref{tab:attack} reports the Area Under Curve (AUC) and the Equal Error Rate (EER) for the ROC curves associated to MIA attacks under different CL configurations. It is possible to notice in many \texttt{Std} set-ups the attack is successful (error lower than $40$ \%), and the adoption of random replay somehow mitigates the inference since more data are processed by the network, and therefore, overfitting is somehow reduced.

On the other side,  the proposed solution (CDML) permits improving the situation since EER values are lifted up to uncertainty ($50$ \%) for all the CDML solutions. Indeed, Table~\ref{tab:attack} shows that the CDML solution obtained the best (dark blue cells) or the second best (light blue cells) performance among all the possible configuration on each task. Notice that EER is always higher than 48 \% in all configurations. The table also reports the value of \texttt{Adapt} approach as well. 
Notice this second strategy works as well, but significant parameter transmission is required. Indeed, weights and biases $\gamma_k , \beta_k$ require the additional transmission of $16384$ bytes for each task $k$, while CDML approach requires communicating $4$ bytes only (i.e., the size of seed value for sequence generation). 

{\bf Ablation studies}\newline
In order to verify the impact of replaying, different runs were performed varying the percentage of data re-processed from the previous tasks. Fig.~\ref{fig:acc} shows accuracy values obtained with $10$ \%, $30$ \%, and $50$ \% of data replayed, while Table~\ref{tab:attack} reports values with no replay, $10$ \%, and $30$ \% of replay data: as the percentage increases, vulnerability slightly decreases since more data are processed by the network. As for modulating solutions (\texttt{CDML},\texttt{Adapt}), high percentages do not change either accuracy or attack success percentages significantly, and therefore, replay can therefore be constrained.

\section{Discussion and final conclusions}

The paper presented a continual learning strategy for class-incremental gait identification systems where embedding features are modulated by means of a Code Division Modulation Layer (CDML). This unit scrambles the generated features according to multiple pseudo-random binary sequences associated to each task. In this way, it is possible to mitigate catastrophic forgetting since latent representations are more uniformly distributed in the feature space. This advantage can be obtained with null or minimal replay thus minimizing the computational cost and privacy leakage. Moreover, the use of a modulating code mitigates membership inference attacks under the considered black-box setting since attackers need to know in advance the associated code. At the same time, it is possible to enable different levels of privacy since different users may be enabled in the recognition of different subsets of enrolled subjects.

Future investigations will be devoted to testing more complex attacks and verifying the accuracy of the protection mechanism on other biometric traces (e.g., faces or irises) and using more complex identification networks (like transformer-based detector, or systems using low-rank adaptation to combat catastrophic forgetting). Moreover, experiments will evaluate the effectiveness of the approach in a dataset-incremental continual learning scheme, where acquisition conditions and the characteristics of samples may vary.

\IEEEtriggeratref{8}
\bibliographystyle{IEEEtran}
\bibliography{refs}

\end{document}